# Application of the metric data analysis method to social development indicators analysis[1] [2]

Kamenev George K.[3], Kamenev Ivan G.[4]

**Abstract:** The article contains a methodology for social statistics assessing. The significance of minorities (groups that differ in their attributes from the majority) has grown substantially in the modern postindustrial economy and society. In the multidimensional characteristics space distribution analysis subjects that not included in the sample can be negligible, but they may be metrically significant. For example, they can be located compactly and remotely from the main mass of subjects, i.e. have similar characteristics and significantly influence the dynamics of socio-economic systems. In addition, it is necessary to evaluate not the probability of errors in each characteristic separately, but the probability of errors in their combinations. The projection of a multidimensional space into two-dimensional or three-dimensional space, usually used to analyze such data, leads to the loss of information about compact minorities that merge into a "false majority" or the appearance of distribution. Thus, the posed task can not be reliably solved without analysis in a multidimensional space. This problem is especially relevant in states aggregated social statistics analysis for which relatively small number of subjects (countries) are represented. The aim of the article is to develop tools for the metric analysis of aggregated social statistics, which is characterized by a small number of measurements. It is developed to check the existence of stable metric and topology structures (clusters) and distribution of countries based on UN data on countries social development characteristics.

## 1 Introduction

These days many theoretical and applied tasks of social sciences are reduced to the analysis of multidimensional sets. In general, they can be described in terms of characteristics collection of different subjects: consumers, voters, workers, enterprises, communities, etc. Subjects are divided into groups (classes) according to combinations of characteristics. Each group is ascribed a certain pattern of behavior. Consumer behavior, models of strategic management, the attributes of human capital, political preferences can be investigated this way. Since the possibilities for collecting mass statistics are limited, we use sample surveys and big data to construct models.

While collecting data on several characteristics of the subject, it is necessary to evaluate not the probability of error in accordance with each characteristic separately, but the probability of errors in accordance with a combination of characteristics. Thus, the task posed cannot be reliably solved without using analysis in a multidimensional space. Moreover, in conditions of the modern postindustrial economy and society, the significance of minorities, that is the groups that differ in their attributes from the majority, has grown substantially. This problem was first described by E. Toffler (Toffler, 1990) and received general recognition in recent years. While analyzing the distribution in a multidimensional space of characteristics, subjects that are not included in the sample are negligible, but may be metrically significant. For example, they can be located compactly and remotely from the bulk of objects, i.e. have similar characteristics and significantly influence the dynamics of socio-economic systems. Disregard of the minorities while constructing the models leads to a significant discrepancy between the actual data and the results of modeling. Thus, it becomes urgent to develop new mathematical methods for the metric analysis of social data in a multidimensional space that would considered both frequency, i.e. measure, and remoteness, i.e. metrics. A clean demonstration of this problem appears in social development indicators analysis, where a small number of countries is characterized by multidimensional indicators' space.

---

[1] This is an extended version of the paper appeared in Conference proceedings of VII International Conference "New trends, strategies and structural changes in emerging markets" (Moscow, May 29-31, 2018).
[2] *The work was prepared with the support of the RFBR grant: 18-01-00465 a: Development of social data multidimensional metric analysis methods*

[3] *Kamenev G.K.*. Government budget institution Federal research center of Informatics and Management Dorodnicyn Computing centre of the Russian science academy. Moscow, Russian Federation. gkk@ccas.ru
[4] *Kamenev I.G.*. Government budget institution Federal research center of Informatics and Management Dorodnicyn Computing centre of the Russian science academy. Moscow, Russian Federation. gkk@ccas.ru

In this investigation, we use the Method of Metric Data Analysis, MMDA, (Kamenev G.K. and Kamenev I.G, 2017 and 2018). It includes the collection of mathematical methods, based on which we created applied mathematical tools and a set of hardware and software designed to analyze data in the multidimensional space of social characteristics attributed to social subjects (humans). An integral part of MMDA for high-dimensional data research is the Data Metric Association (DMA) method, which consists in multidimensional data set metric neighborhood connected components construction and investigation.

## 2 Social development indicators usage conceptualization

The UN database on various human development indices (Human Development Data) contains data for 1990-2017 years. The characterized subjects in this case are countries. Each indicator is a quantitative aggregate summarizing the country state in one or another social life aspect. Let us select a number of indicators characterizing human capital and potential:

1. Expected years of schooling (years)
2. Employment to population ratio (% ages 15 and older)
3. Education Index
4. Inequality in income (%)
5. Inequality in life expectancy (%)
6. Internet users (% of population)
7. Life expectancy at birth (years)
8. Public health expenditure (% of GDP)

Together, these data form an 8-dimensional characteristics space, to which can be added the 9th dimension, the year. However, the scale of the problem does not allow us to report research methodology and data dynamic regularities in one publication, so it was decided to choose a static 8-dimensional space.

The most complete data is available for 2015 (with the exception of Public health expenditure, for which were used data for 2014). The whole selected indicators (characteristics of countries) values set is present for 151 countries from the general list, which includes 195 countries. These 151 countries were included in the investigated data set, for which topology (spatial structure) and metric study task was solved.

## 3 Multidimensional social data analysis methodology

The multidimensional space of characteristics is a set of characteristics simultaneously attributed to the subject.

In particular, it can be a space of social characteristics (education, health, etc.). The general principle of this approach is that the laws of behavior of a subject are determined by a combination of its studied characteristics. Each characteristic should be not qualitative, but quantitative, i.e. to assume a certain method of quantitative measurement. The next step is the analysis of subjects position in space, their typification and classification.

Attempts to classify subjects and describe the characteristics of their behavior are often undertaken in various social sciences. In particular, this applies to categories and types, allocated on the basis of sets of characteristics and multi-level classifications. In view of the insufficient prevalence of methods of quantitative multivariate analysis, these models are often artificially transformed into qualitative ones. In particular, subjects and institutions can be grouped into two opposite categories (for example, "developed" and "developing" countries)

However, such models need a serious empirical justification, which, unfortunately, often turns out to be "upside down": it justifies the correspondence of specific subjects or institutions to previously allocated categories, rather than the very existence of these categories.

Application of the metric data analysis method to social development indicators analysis

Rapid evolution of social systems requires the creation of techniques that allow us to revise categorization directly in the process of research and adjust the allocated classes. Accordingly, factor analysis should be supplemented by the analysis of subjects distribution in space and grouping them into structures.

It is proposed to implement the study of multidimensional space of characteristics, following a specific algorithm of actions.

1. Data collection on available characteristics.
2. Investigation characteristics selection.
3. Construction of multidimensional space.
4. Explicit construction (approximation) of the general population.
5. Structural and visual analysis.
6. Generalization of characteristics and study of the characteristics of the behavior of the selected classes.

While locating observation objects in a multidimensional space, it is necessary to distinguish types, classes, isolated groups based on spatial (metric) proximity. Visualization, cluster analysis, structural analysis, various applied mathematical methods, including the construction of projections into spaces of smaller dimension, can be applied to solve this task. Allocated groups of subjects should be described in detail, having selected key characteristics, which similar values determines their association. In the future, a deeper factor analysis of each class and a description of the characteristics of its functioning become possible.

## 4 Metric data neighborhood topology

The set in topological space is called connected if it cannot be represented as the union of two or more disjoint nonempty closed (or open) subsets. It means that the connected set consists «of one piece». Union of all connected subsets containing some point is called the connected component of this point.

Let there be data in a numerical format – a finite D of cardinality (number of elements) $N$ in $d$-dimensional Euclidean space. We consider a topological space with domain $D(E)$ – the closed metric $E$-neighborhood of $D$ (points of space on distance from $D$ that is less or equal to $E$), and the topology of the closed metric neighborhoods of its points. The $E$ value we called the *association threshold*. Then the two data points will belong to the same connected component if the distance between them is less or equal to $2E$. Therefore the two data points will belong to the same connected component if there is a polygonal chain of connected data points (pairs of points intersecting with their closed $E$-neighborhoods) forming a continuous polygonal line joining the set points and located completely in data $E$-neighborhood $D(E)$.

Let us consider the graph with data points as vertices connected by edges, if their closed metric $E$-neighborhoods intersects. We obtain data graph which corresponding to this threshold association. Then designate it as $GA(E)$ and call it *the data metric association graph*. In graph theory, a graph is connected when there is a path between every pair of vertices, and *the connected component* is a maximum connected subgraph. It is easy to see that the concepts of the connected components of considered topological space $D(E)$ and components of the association graph $GA(E)$ coincides. In the future, we will only mention the association graph.

Let's denote by $K(E)$ *the connectivity number* (number of connected components) of the data metric association graph $GA(E)$. The ratio $k(E) = K(E)/N$ we call *the data metric dissociation coefficient*. When $E = 0$, if there is no matching data points, $k(E) = 1$. Then, there is value $R$ for which when $E$ is greater or equal to $R$ it follows $k(E) = 1 / N$. Therefore, it is clear that the chart of $k(E)$ is a monotonically decreasing function.

The dependence of $k(E)$ from the association threshold $E$ we call *the data metric association chart*. Sharp drops of this chart means the topological data metric neighborhoods consolidation in clusters with various degree complexity of internal structure.

Kamenev G.K., Kamenev I.G.

## 5 United Nations human development indicators data metric analysis

Let us consider the entire collection of data as a set of points in an 8-dimensional Euclidean space, and let us choose the relative Chebyshev distance as a metric. In this case, the distance between two points is the maximum difference of points among the components multiplied by the weights proportional to the available ranges of the data set. The distance between countries reflects their similarity, and the division of countries into any classes (such as "developed" and "developing" countries) should reflect their topological closeness. To put it more precisely, in order to divide countries into classes it is necessary to have a clear boundary between them. It means that there should be no topological connection between the corresponding clusters. Surely, structures can also be identified within the cluster by methods of density analysis, but this is a different problem that deserves a special investigation.

As can be seen, the distance between the subjects (countries) is the criterion of connectivity. That is why the set of countries can be represented as $E$-net, and the allocated clusters as $E$-covering. This allows applying MMDA for topology analysis by comparing the coverage form for different $E$.

The specific of the initial data array can be high consolidation or, on the opposite, high diffusivity. For example, in the studied case a large number of countries form single clusters that are not inclined to join any groups. They cease to be single only after all the clusters merge into the big one. Such single subjects may be as a rule excluded from investigation, and only clusters that consist of two or more countries will be of general interest.

## 6 UN countries data association and dissociation research

Let us carry out a systematic metric analysis and analyze the UN data on various indices of human development (Human Development Data, 2015). The objects of observation in this case are countries. Let us distinguish a number of indicators related to human capital. The most complete data are available for 2015. It is required to characterize metrically the whole set of countries in accordance with the selected indicators and, if possible, to identify substantial structures in it.

Let us consider data for 151 countries in 8 characteristics:

1) Expected years of schooling (years) < ExpYSch >
2) Employment to population ratio (% ages 15 and older) < EmplPRatio >
3) Education Index < EduInd >
4) Inequality in income (%) < IneqIncome >
5) Inequality in life expectancy (%) < IneqLifeExp >
6) Internet users (% of population) < InterUsers >
7) Life expectancy at birth (years) < LifeExpBirth >
8) 2014 Public health expenditure (% of GDP) < PubHealExp >

Here the angle brackets indicate a brief characteristic naming.

Let us investigate the metric association of these data in multidimensional clusters. In the Fig. 1, we can see a DMA chart. The value of $E$ measures the metric association threshold (the size of the metric neighborhood of the data set) and is the relative distance in the Chebyshev metric.

At the beginning of the chart there is an area we call *the primary metric data association domain*. In this area, at $E = 0.024$ (2.4% of the range of indices variation) in the diffuse cloud of the metric data neighborhoods, the first connected component is condensed, containing two countries: {Canada, UK}. If we look further ($E = 0.028$), we can see the clusters formed, which contain groups of countries similar in their development. Characteristically, most often these are neighbor countries. A sharp consolidation occurs at $E = 0.06$, when a large cluster of developed countries is formed {Norway, Switzerland, Germany, Denmark, Netherlands, Canada, Great Britain, Japan, France, Belgium, Finland, Austria, Slovenia, Czech Republic}. At $E = 0.072$ we can see the end of substantial clusters formation (see below). We will call this middle area of the chart *the substantial data metric association domain*.

At $E = 0.1$ the process of association of the main data set takes place. Different marginal stay outside of it, for example diffuse clusters (which consists only of one data point) {Azerbaijan}, {Maldives}, a small

Application of the metric data analysis method to social development indicators analysis cluster {Gabon, Iraq, Timor-Leste, Syrian Arab Republic} and diffuse clusters of weakly developed African countries. At the end of the chart, we can see *the primary data metric dissociation domain*: at $E = 0.164$ {Maldives} separate from the single cluster of all countries, at $E = 0.16$ the data are dissociated by clusters {Swaziland, Lesotho}, {Nigeria}, {Djibouti} and {Central African Republic}.

Let us now investigate one of the intrinsic values of the association threshold, lying in the area of substantial metric association domain: the point $E = 0.072$ (7.2% of the range of indices variation) highlighted with red in Fig. 1.

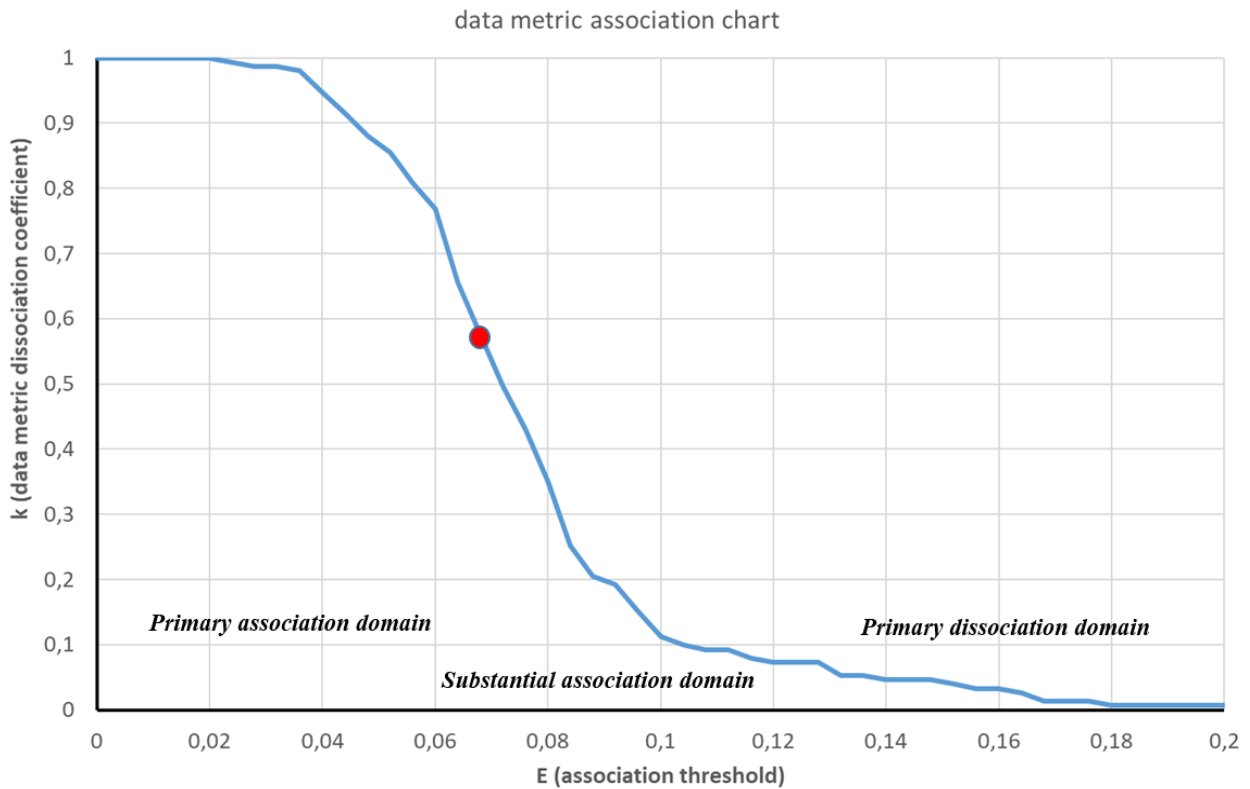

Fig. 1. The data metric association chart

For this value of association threshold $E$ all 8-dimensional data associates in 75 connected components of data metric $E$-neighborhood with 60 diffuse clusters (containing isolated neighborhoods of single data point) and 15 condensed clusters for which are of interest:

- 1st cluster, containing countries with relative high social development – {Norway, Australia, Switzerland, Germany, Denmark, Netherlands, Ireland, Iceland, Canada, Sweden, United Kingdom, Japan, Korea (Republic of), France, Belgium, Finland, Austria, Slovenia, Italy, Spain, Czech Republic, Greece, Estonia, Cyprus, Malta, Poland, Lithuania, Slovakia, Portugal, Hungary, Latvia, Croatia, Montenegro, Russian Federation, Romania, Belarus, Bulgaria, Kazakhstan, Serbia, Turkey, Albania, Ukraine, Jordan, Tunisia, Moldova (Republic of)}.
- 2nd cluster – {United States}, this is a diffuse (single) cluster of special interest. Amazing feature of {US} cluster consist in its stabile diffusion property: this cluster (country) joins to the 1st cluster of relative high social development countries only at $E=0.1$. At this association threshold value this is strictly speaking "developed and developing" country's cluster (with marginal excluded). This shows the illegality of US association with the rest of the "developed countries", the presence of their fundamental features of social development.
- 8th cluster, containing countries with relative week social development – {Panama, Costa Rica, Venezuela (Bolivarian Republic of), Mexico, Brazil, Peru, Thailand, Ecuador, China, Saint Lucia, Jamaica, Colombia, Dominican Republic, Paraguay}. A distinctive feature of this cluster is the presence in it a big number of Latin American countries, which therefore constitute a special type of social development. The fact of China belonging to this cluster is of special interest.
- 46th cluster, containing the least social-developed countries – {Cambodia, Nepal, Tanzania (United Republic of), Madagascar}.

Kamenev G.K., Kamenev I.G.

Figg. 2, 3 and 4 show countries profiles (plotted for 8 indicators) of the respective clusters (except 2nd). Here, the values of each parameter are normalized by the maximum of it in the population. It is clearly seen that profiles of all the countries in the same cluster are similar in shape, but differ significantly for different clusters. Fig. 5 shows the single country profile of the 2nd cluster (United States, black) on the background of countries group forming the 1st cluster (brown palette). We can see why this country is not included in this cluster – it stands out by "Inequality in income" value, which in this country is much higher than in the countries of the 1st cluster (the absolute value of 27 compared with 9.2-21.8).

Fig. 6 shows a three-dimensional projection of data metric $E$-neighborhood ($E = 0.072$) where data subspace contain three indicators: EduInd (abscissa or $x$-axis), IneqIncome (ordinate or $y$-axis), PubHealExp (applicate or $z$-axis). On that figure cubes collection of each color corresponds to one of the above eight-dimensional topologically connected metric components: red color corresponds to the 1st cluster, white – 2nd cluster, green – 8-th cluster, blue – 46-th cluster, gray color cubes are presenting metric neighborhood of other data. We can see a well distinguishable specificity in metric localization of selected components, they even do not intersect with each other despite of the fact that the disconnection in the original eight-dimensional space does not guarantee the disconnection in the three-dimensional projection. In this case, however, they are not penetrated one in another how it is done by many diffuse clusters, which can not be attributed to the major clusters mentioned above, due to significant deviations in other non-projection indicators. This fact illustrates the danger of using a fundamentally different low-dimensional projections and a quality of the results obtained by multivariate analysis (in this case, by MMDA).

## 7  Results

Method of Metric Data Analysis allowed to analyze the subjects' general population representation by existing social data block, and reveal its topology components and metric distribution for multiple characteristics. The revealed topological structures qualitative characteristics were given from the point of view of internal countries social policy. We conclude that the countries grouping in terms on the social development level is more complex than the common classification into developed and developing ones. In particular, the Latin American countries differ from the majority of developing countries by the higher expected years of schooling, the education index and income inequality. Moreover, the United States of America has a different model of social development than the other developed countries, due to high income inequality (which does not affect other indicators of social development). It is shown that a large number of countries (like the USA) do not fit into the main clusters and therefore need special study. The prospects for using MMDA for analyzing the stability of clusters, dynamics of their boundaries and composition from year to year deserve special mention. These questions are the subject of an independent research.

Application of the metric data analysis method to social development indicators analysis

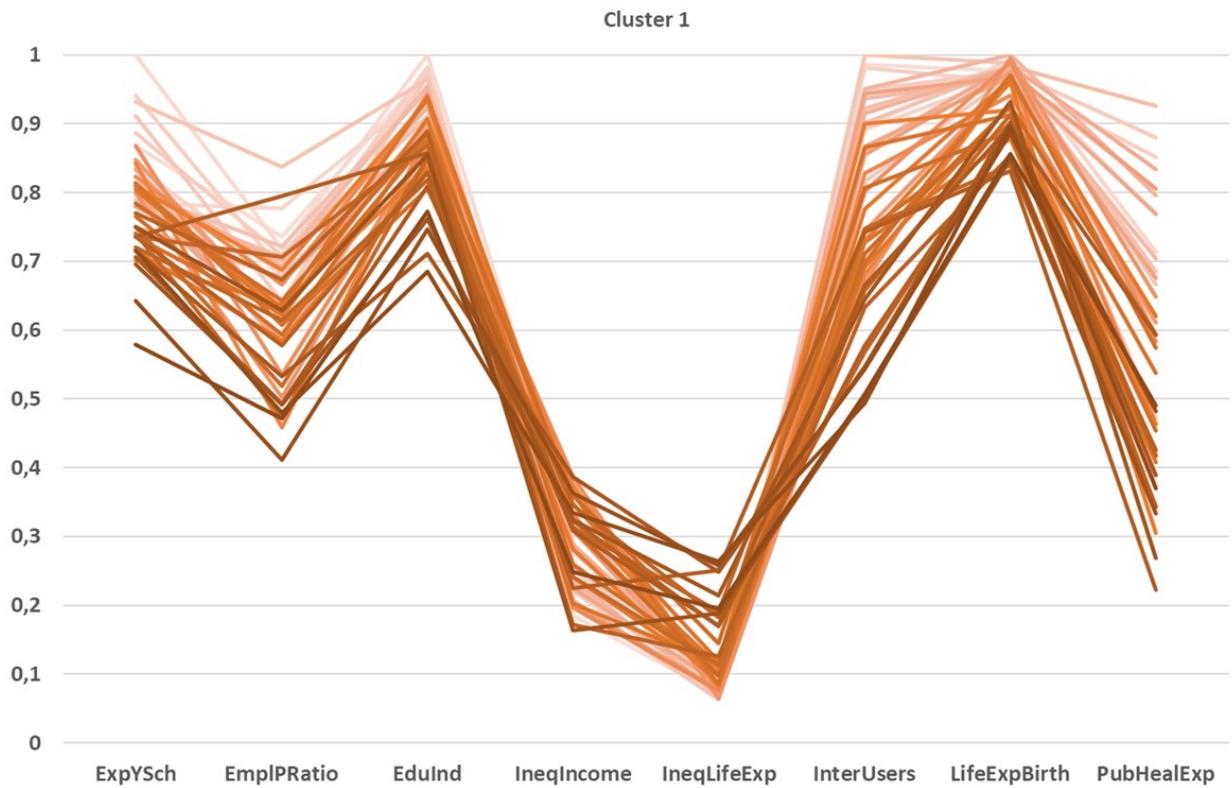

Fig. 2. Profiles for 1st cluster countries (relative high social development)

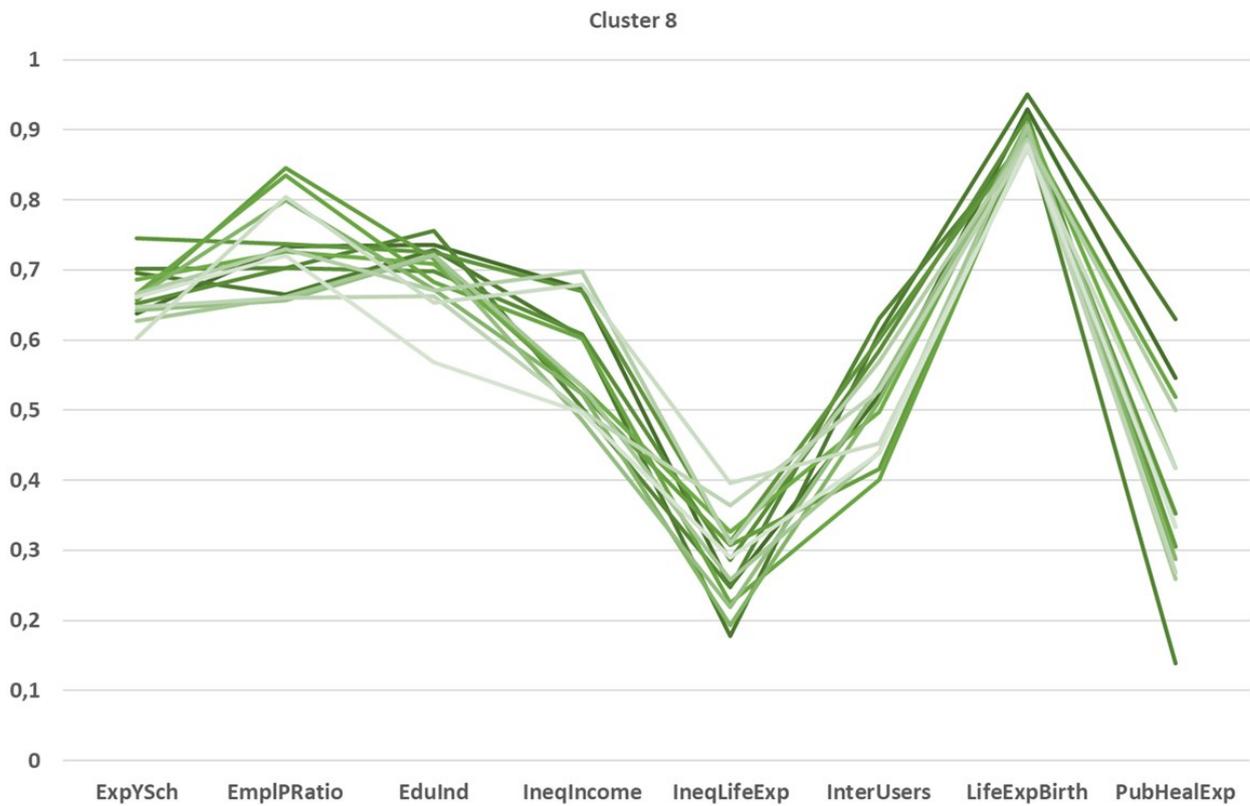

Fig. 3. Profiles for cluster 8th countries (relative week social development)

Kamenev G.K., Kamenev I.G.

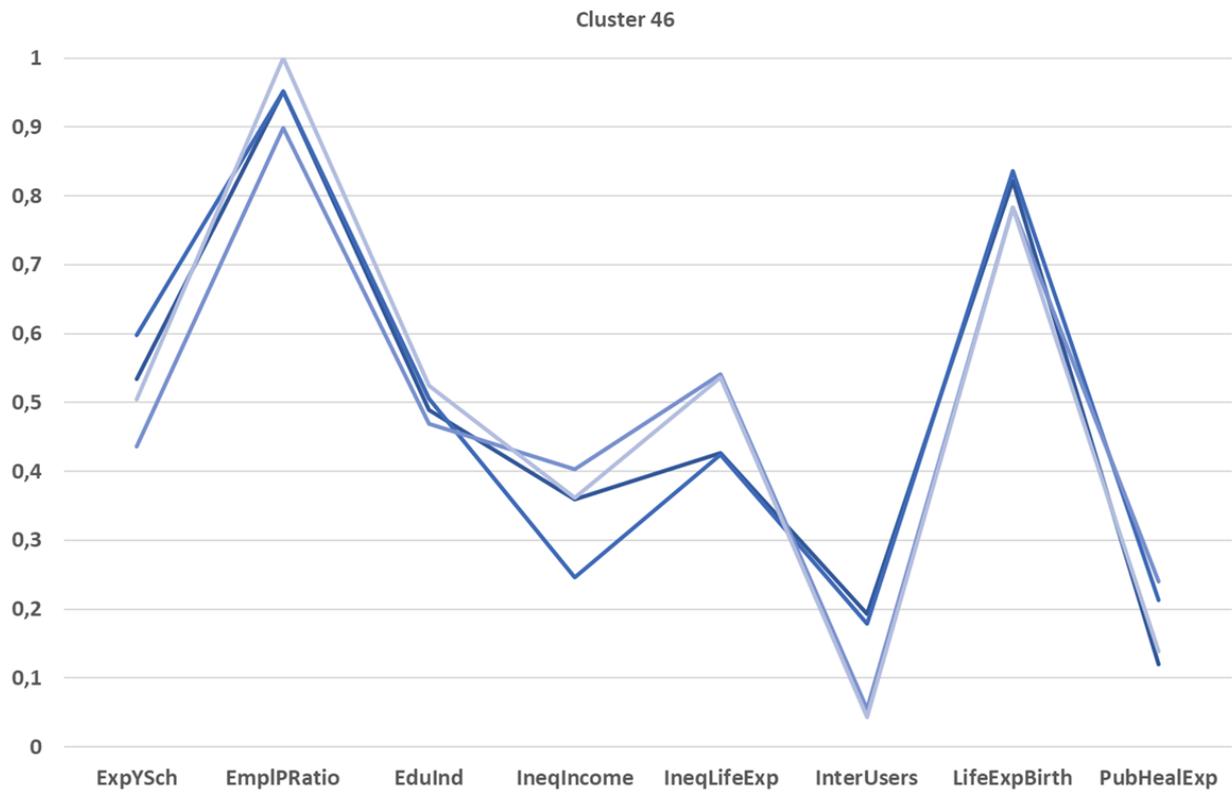

Fig. 4. Profiles for 46th cluster countries (the least social-developed countries)

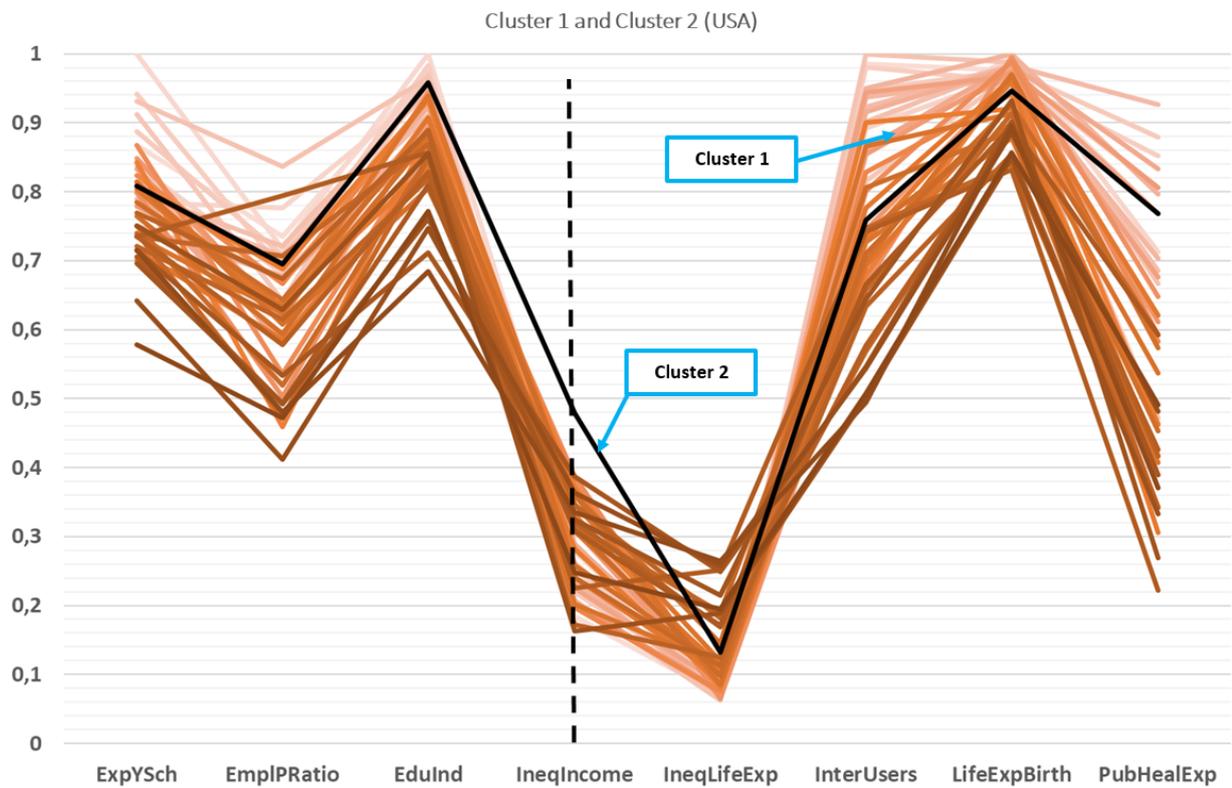

Fig. 5. Profile for 2nd cluster state (USA) on the background of 1st cluster countries profiles

Application of the metric data analysis method to social development indicators analysis

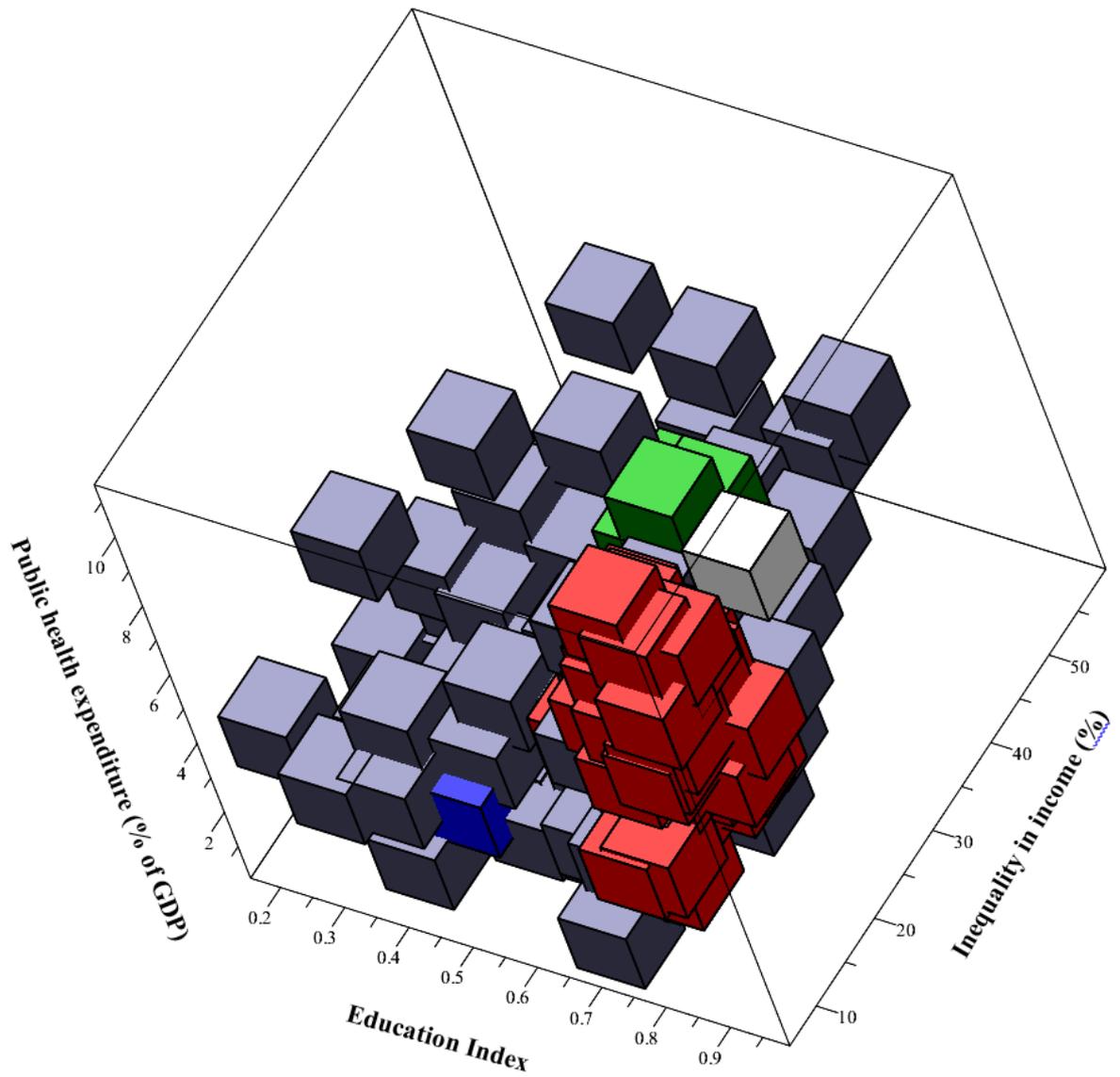

Fig. 6. The 3-dimensional projection of 8-dimensional data metric neighborhood: cluster 1 (red), cluster 2 (white), cluster 8 (green), cluster 46 (blue)